\documentstyle[prl,aps,multicol,epsf]{revtex}
\begin{document}
\draft
\widetext

\title{Theory of Underdoped Cuprates}

\author{Xiao-Gang Wen, and Patrick A. Lee }
\address{Department of Physics, Massachusetts Institute of
Technology,  Cambridge, Massachusetts 02139}

\date{June, 1995}
\maketitle

\widetext
\begin{abstract}
\leftskip 54.8pt
\rightskip 54.8pt
We develop a slave-boson theory for the $t$-$J$ model at finite doping
which
respects an $SU(2)$ symmetry -- a symmetry previously known to
be important at half filling.
The  mean field phase diagram is found to be consistent
with the phases observed in the cuprate superconductors, which contains
$d$-wave superconductor, spin gap, strange metal, and Fermi liquid
phases. The spin gap phase is best understood as the staggered flux
phase,
which is nevertheless translationally invariant for physical quantities.
The  electron spectral function shows small Fermi pockets
at low doping which continuously evolve into the large Fermi
surface at high doping concentrations.
\end{abstract}


\begin{multicols}{2}

\narrowtext
The parent compound of the cuprate superconductors is an
antiferromagnetic
(AF) insulator.  With hole doping, AF is rapidly destroyed and a
metallic
state emerges.  It is well established that at optimal doping, a Fermi
surface exists with area $1 - x$ where $x$ is the concentration of
doped holes.\cite{1}  On the other hand, for
$x \ll 1$, the system remains AF with a doubled unit cell, and the
Fermi
surface is expected to be small pockets centered around the
$({\pi\over 2}, {\pi\over 2})$ point.\cite{2}
An important question is how the low-lying electron state evolves from
small $x$ to optimal doping.  This intermediate region, called the
underdoped regime, also exhibits unusual magnetic properties often
referred
to as the spin gap behavior.  Unlike optimally doped systems, where the
magnetic susceptibility $\chi$ is temperature independent, underdoped
cuprates generally show a reduction in $\chi$ at temperatures below
400K or so.\cite{3}
  Below 150K, $\chi$ and the $NMR$ relaxation rate decreases
abruptly
in an activated manner.  It has been argued that this is observed only
in
bilayer material.\cite{3}
In this paper we shall concentrate only on the high temperature
spin gap behavior, which we view as evidence for the formation of spin
singlets within the $Cu$-$O$ layer.  We shall address the question of
how the spin gap manifests itself in the electronic spectral function,
 and how the Fermi
surface evolves from small pockets to a large Fermi surface which
satisfies
Luttinger's theorem.  This last question was addressed in a weak
coupling
theory\cite{4} where fluctuating spin density waves induce shadow
bands.
  We would like to study the strong correlation limit
 which we believe to be more
appropriate for the cuprates, and we take the $t$-$J$ model as our
starting
point.

A standard way of enforcing the constraint of no double occupancy of the
$t$-$J$ model is to write the electron operator
$c_{\alpha i}$ in terms of auxiliary fermions and boson particles
$c_{\alpha i} = f_{\alpha i} b_i^\dagger$
and demanding that each site is occupied by either a fermion or a boson.
In a mean field (MF) treatment, the order parameters
$\chi_{ij} = \langle f_{\alpha i}^\dagger f_{\alpha j}\rangle$ and
$\Delta_{ij}=\langle f_{1i}f_{2j}
-f_{2i}f_{1j}\rangle$
describes the formation of singlets envisioned in Anderson's resonating
valence bond (RVB) picture.\cite{7}
At zero doping, the translationally invariant solution can be described
as a
$\pi$-flux phase\cite{8}
or a $d$-wave pairing
state with $|\Delta_{ij}| = |\chi_{ij}|$.
The symmetry which underlies the degenerate MF states has
been identified as $SU (2)$, which expresses the idea that a
physical up spin can be viewed as either the presence of an up spin or
the absence of a down spin fermion.\cite{9}  In the conventional
theory, the
$SU (2)$ is broken to $U(1)$ upon doping, and only the $d$-wave state
survives as the MF solution.\cite{10,11}  This scenario has been used
as
an
explanation of the spin gap phenomenon.\cite{11}

In this paper we present a new formulation of the constraint which
preserves
$SU(2)$ symmetry away from half-filling.  Our hope is that since $SU(2)$
is an
exact symmetry at half-filling, the MF approximation of the new
formulation may capture more accurately the low energy degrees of
freedom
and may be a better starting point for small $x$.  We are also motivated
by
the photo-emission experiment on the insulating cuprate,\cite{12} which
finds
a large excitation energy at the $(0, \pi )$ point,
comparable
to that at $(0, 0)$.  This is just what is expected from the
$\pi$-flux phase spectrum, suggesting that the AF state may resemble
the
$\pi$-flux phase at short distances.  As we shall see, the $SU(2)$
formulation
provides a scenario for how the $\pi$-flux phase is connected to
the spin gap phase and how the hole
pockets evolve upon doping.

The $SU(2)$ doublets
$\psi_{1i} =  \left ( {f_{1i}\atop f^\dagger_{2i}} \right ) $ and
$\psi_{2i} = \left ({f_{2i}\atop -f^\dagger_{1i}}\right )$
were introduced in Ref.\cite{9}.
Here we introduce {\it two} spin-0 boson fields
$b_a$, $a = 1, 2$ forming another doublet
 $b_i = \left ({b_{1i}\atop b_{2i}}\right ) $.
We then form $SU(2)$ singlets to represent the following
physical operators
$\vec S_i = {1\over 2}
f^\dagger_{\alpha i} \vec\sigma_{\alpha \beta} f_{\beta i}$,
$c_{1i}
=  b^\dagger_i\psi_{1i} /\sqrt 2
=   ( b^\dagger_{1i} f_{1i}
+ b^\dagger_{2i}f^\dagger_{2i} ) /\sqrt 2
$, and
$c_{2i}
=  b^\dagger_i \psi_{2i} /\sqrt 2
=  ( b^\dagger_{1i}f_{2i}
- b^\dagger_{2i}f^\dagger_{1i} ) /\sqrt 2
$.
The $t$-$J$ Hamiltonian
$\sum_{(ij)}  [ J  (\vec S_i\cdot \vec S_j - {1\over 4} n_i
n_j
 ) -t(c^\dagger_{\alpha i} c_{\alpha j} + h.c.) ]$
can now be written in terms of our fermion-boson (FB) fields.
The Hilbert space of the FB system is larger than that of the
$t$-$J$ model.  However, the local $SU(2)$ singlets satisfying
$ ({1\over 2} \psi^\dagger_{\alpha i} \vec\tau \psi_{\alpha i}
+ b^\dagger_i \vec\tau b_i  ) |{\rm phys} \rangle = 0$
form a subspace that is identical to the Hilbert space of the $t$-$J$
model.
On a given site, there are only three states that satisfy the
above constraint.  They are
$f^\dagger_1 |0\rangle$, $f^\dagger_2 |0\rangle$, and
${1\over \sqrt 2}  (b^\dagger_1
+ b^\dagger_2 f^\dagger_2 f^\dagger_1  ) |0\rangle$
corresponding to a spin up and down electron, and a vacancy
respectively.
Furthermore, the FB Hamiltonian,
 as a $SU(2)$ singlet operator, acts within the subspace, and has
same matrix elements as the $t$-$J$ Hamiltonian.

Following the standard approach, we
obtain the following MF Hamiltonian\cite{UL}  for the FB system:
\begin{eqnarray}
&&H_{\rm m}  =
 - \mu \sum_i b^\dagger_ib_i
-\sum_i a^l_{0i}  ({1\over 2} \psi^\dagger_{\alpha i}
 \tau^l\psi_{\alpha i} + b^\dagger_i \tau^l b_i) +\cr
&& \sum_{(i, j)} {3J\over 8}
( |\chi_{ij} |^2 + |\Delta_{ij}|^2
+ \psi^\dagger_{\alpha i} U_{ij} \psi_{\alpha j}) +
t (b^\dagger_i U_{ij} b_j + {\rm h.c.} )
\label{5}
\end{eqnarray}
where
$U_{ij} = \left ( {-\chi^*_{ij}\atop \Delta^*_{ij}}\quad
{\Delta_{ij}\atop \chi_{ij}}\right )$.

The first two terms of $H_{\rm m}$ are included to impose the
constraints.  The
value of $\mu$ is chosen such that the total boson density
(which is also the density of the holes in the $t$-$J$ model) is
$\langle b^\dagger_i b_i\rangle =
\langle b^\dagger_{1i} b_{1i} +  b^\dagger_{2i} b_{2i}\rangle = x$.  The
values of $a^l_{0i}$ are chosen such that
$\langle {1\over 2} \psi^\dagger_{\alpha i} \tau^l \psi_{\alpha i} +
b^\dagger_i \tau^l b_i\rangle = 0$.
For $l = 3$ we have
\begin{equation}
\langle f^\dagger_{\alpha i} f_{\alpha i} +
b^\dagger_{1i} b_{1i} - b^\dagger_{2i} b_{2i}\rangle = 1 \label{6}
\end{equation}
We see that unlike the $U(1)$ case the density of the fermion
$\langle f^\dagger_{\alpha i} f_{\alpha i}\rangle$
is not necessarily
equal $1 - x$.  This is because a vacancy in the $t$-$J$ model
may be
represented by an empty site with a $b_1$ boson, or a doubly occupied
site with a $b_2$ boson.  We also notice that the MF Hamiltonian is
invariant
under local $SU(2)$ transformations,
$W_i \in SU(2)$:
$\psi_{\alpha i} \to W_i \psi_{\alpha i}$,
$b_i \to W_ib_i$,
$U_{ij} \to W_i U_{ij}W^\dagger_j$, and
$a^l_{0i}\tau^l \to W_i a^l_{0i} \tau^l W_i^\dagger$.

We have searched the minima of the MF free energy for the
MF
ansatz with translation, lattice and spin rotation
symmetries.
We find a phase diagram with six different phases (Fig. 1).

{(1)} Staggered flux (sF) phase:
\begin{equation}
\cases{
U_{i, i + \hat x} & = $- \tau^3 \chi - i (-)^{i_x+i_y} \Delta$\cr
U_{i, i + \hat y} & = $- \tau^3 \chi + i (-)^{i_x+i_y} \Delta$\cr
}\label{7}
\end{equation}
and $a^l_{0i}  = 0$.
In the $U(1)$ slave-boson theory, the staggered flux phase breaks
translation symmetry.
Here the breaking of translational invariance is a gauge
artifact.
In fact, a site dependent $SU(2)$ transformation
$W_i = {\rm exp} ( i (-1)^{i_x + i_y} {\pi\over 4} \tau_1 )$
 maps the sF
phase to the $d$-wave pairing phase of the fermions:
$U_{i, i+ \hat x,\hat y} = -\chi\tau_3 \pm \Delta\tau_1$,
which is explicitly translationally invariant.
In the sF phase
 the fermion and boson dispersion are given by
$\pm E_f$ and $\pm E_b$, where $E_f =
 \sqrt{(\epsilon_f-a_0^3)^2 + \eta^{2}_f}$, $\epsilon_f  =
-{3J\over 4} (\cos k_x + \cos k_y ) \chi$,
$\eta_f  = -{3J\over 4} (\cos k_x - \cos k_y ) \Delta$,
and a similar result for $E_b$ with ${3J\over 4}$ replaced by $2t$.
Since $a^3_0 = 0$ we have
$\langle f^\dagger_{\alpha i} f_{\alpha i} \rangle = 1$ and
$\langle b^\dagger_1b_1\rangle =
\langle b^\dagger_2b_2\rangle = {x\over 2}$.

{(2)}  The $\pi$-flux ($\pi F$) phase is the same as the sF phase
except
here $\chi = \Delta$.

{(3)}  The uniform RVB (uRVB) phase is described by Eq. (\ref{7})
with $a_{0i}^l=\Delta = 0$.

{(4)}  A localized spin (LS) phase has
$U_{ij} = 0$ and $a^l_{0i} = 0$, where
the fermions cannot hop.
\begin{figure}
\epsfxsize=3.5truein
\centerline{\epsffile{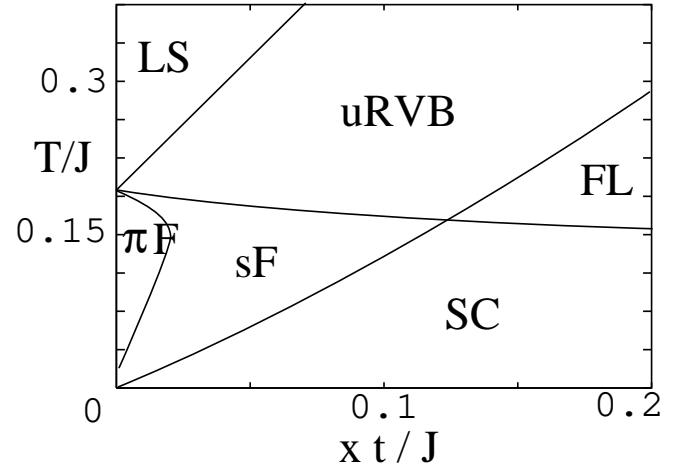}}
\caption{
$SU(2)$ MF phase diagram for $t/J=1$. The phase diagram for $t/J=2$
is quantitatively very similar to the $t/J=1$ phase diagram, when
plotted in terms of the scaled variable $xt/J$, except the
$\pi$F phase disappears at a lower scaled doping concentration.
}
\label{fig1}
\end{figure}

{(5)}  The $d$-wave superconducting (SC) phase is described by
$U_{i, i+ \hat x,\hat y} = -\chi\tau_3 \pm \Delta\tau_1$
and $a^3_0 \ne 0$, $a^{1, 2}_0 = 0$, $\langle b_1\rangle \ne 0$,
$\langle b_2\rangle = 0$.
Notice that the boson condenses in the SC phase despite the fact that
in
our MF theory the interactions between the bosons are ignored.
The SC MF solution provides an interesting example of
finite-temperature free boson
condensation in two dimensions.
To see this, notice that
the $a^3_0$ term
in the FB Hamiltonian
$-\sum_i a^3_0  ( f^\dagger_{\alpha i} f_{\alpha i} +
b^\dagger_{1i} b_{1i} - b^\dagger_{2i} b_{2i} - 1  ) $
makes $a^3_0$
behaves like the chemical potential of the fermions.  The fermions
favor a non-zero $a^3_0$.
Let us assume $a^3_0 < 0$, which makes
$\langle f^\dagger_{\alpha i} f_{\alpha i}\rangle < 1$.
A negative $a^3_0$ also makes
the $b_1$-band bottom to be higher than that of $b_2$, and
the thermally excited bosons satisfy
$ \langle b^\dagger_1 b_1\rangle_{therm}
< \langle b^\dagger_2 b_2\rangle_{therm}$.
Thus the thermally excited bosons alone cannot satisfy the
constraint in  Eq.(\ref{6}).
The $b_1$ boson are forced to condense at the bottom of the $b_1$ band
to satisfy the constraint, in the same way that ordinary
boson condense to satisfy the density constraint.
Due to the fermion contribution,
the total free energy can still be lowered
by generating a finite
$a^3_0$ at low temperatures.

{(6)}  The Fermi liquid (FL) phase is similar to the SC
phase except that there is no fermion pairing ($\Delta = 0$).

In the following we would like to discuss some simple physical
properties
of the MF phases.  uRVB, sF, $\pi$F and LS phases contain
no boson
condensation and correspond to unusual metallic states.  Since
$a^3_0 = 0$, the area of the fermion Fermi sea in the uRVB phase
is pinned at ${1\over 2}$ of the Brillouin zone.  As
temperature is lowered, fermions condense in pairs, or develop staggered
flux and the uRVB phase changes into the sF or $\pi$F phases.
A gap is opened at the Fermi surface which reduces the low energy spin
excitations.  Thus the sF and $\pi$F phases correspond
to the high temperature  spin gap phase.
  The SC phase contains both the boson
and the fermion-pair condensations and corresponds to a $d$-wave
superconducting
state of the electrons.  The FL phase containing boson
condensation
corresponds to a Fermi liquid phase of electrons.  However, the area
of
the Fermi sea produced by the $SU(2)$ MF theory is larger than
that
predicted by the Luttinger theorem which reveals a drawback of the
$SU(2)$ MF theory at high doping concentrations.  This is
probably
related to another drawback of the $SU(2)$ theory that superconducting
$T_c$ goes down too slowly beyond the optimal doping.  It appears that
the
$U(1)$ MF theory is better at higher doping.

\begin{figure}
\epsfxsize=3.5truein
\centerline{\epsffile{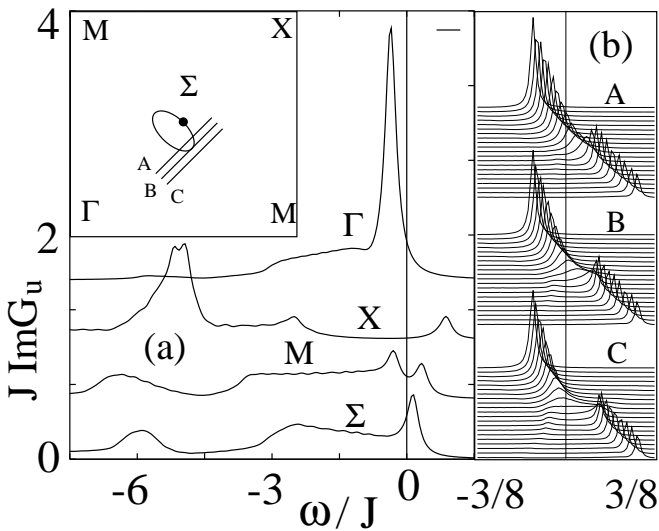}}
\caption{
(a) The electron spectral function Im$G_U$ in the sF phase for
$t/J=2$, $x=0.041$, $T/J=0.13$, where
$\chi=0.57$ and $\Delta=0.22$.
The sharp peaks near $\omega=0$ are quasi-particle peaks.
The insert shows a quarter of the Brillouin
zone.  (b) The spectral functions Im$G_U$
for three linear scans along the line A, B, and C in the insert.
}
\label{fig2}
\end{figure}
We would like to point out that the different MF phases contain
different gauge symmetries.  The uRVB and the $\pi $F phases have the
full
$SU(2)$ gauge symmetry.
In the sF phase
the $SU(2)$ gauge symmetry is broken down to $U(1)$,\cite{gaug}
while in the SC and FL
phases the $SU(2)$ gauge symmetry is completely broken.

Next we calculate the physical electron Green function in the sF phase.
Using the
expression of $c_\alpha$,
the MF approximation $G_0$ is given by the convolution of fermion and
boson Green functions.
The expression of $G_0$ is lengthy, but can be approximated at
low temperatures by
\begin{equation}
G_0 (k, \omega ) = {x\over 2}
\left ( {u^2\over \omega - E_f} + {v^2\over \omega + E_f}\right ) +
G_{in}
\label{13}
\end{equation}
The first term describes the coherent motion of electrons with the
fermion
dispersion. The new feature is the appearance of the coherent factors
$u (k) = \sqrt{{E_f + \epsilon_f\over 2E_f}} {\rm sgn} (\eta_f )$ and
$v (k) = \sqrt{{E_f - \epsilon_f\over 2E_f}} $.
The second term is the incoherent background which mainly reflect the
boson
density of states. Im$G_{in}$ exists only for $\omega<0$ and contributes
$1/2$ to a total spectral weight $(1+x)/2$.

The coherent part of $G_0$ produces only Fermi points at
$\Sigma=(\pm \pi/2,\pm \pi/2)$.
Another feature is that the occupied part of
the
spectral weight of $G_0$ contains $1+{x\over 2}$ electrons as opposed to
$1-x$
electrons. These unsatisfactory features are due to the absence of
correlation
between fermions and bosons in arriving at $G_0$. In reality there is a
strong
attraction between them due to gauge fluctuations. In the limit of a
single
hole, this attraction can lead to a bound state with the quantum number
of a electron,
as emphasized in Ref. \cite{14}. In the case of finite hole
concentration,
we expect that fermion particle-hole pairs may be spontaneously excited
out of the MF ground state so that the $b_2$ ($b_1$) bosons can  bind to
fermions (anti-fermions). The result would be low lying physical
electron excitations which may resemble a Fermi surface. In order to
capture this physics, even at a very crude level, we assume that
after screening, the $a_0^l$ fluctuations induce the following
short range interaction
$- \sum_i U \psi^\dagger_{\alpha i} \tau^l \psi_{\alpha i}\
b^\dagger_i\tau^l b_i$, where $U < 0$.
We have calculated $G_U$ which included this attraction in the
Bethe-Salpeter approximation. The results are shown in Fig. 2 and 3.
To interpret these results, we consider instead
$G'_U=(G_0^{-1}+U)^{-1}$ which we have found numerically to be nearly
identical
to
$G_U$. It is easy to see that $G'_U$ also has the form
$G'_U=G'_{coh}
+G'_{in}$. In the case when either $|u|$ or $|v|\approx 1$,
 we find that
$G'_{coh}={Z\over \omega \pm E_f+\mu_f}$ with
$\mu_f={xU\over 2 (1 + U G_{in})}$ and $Z={x/2\over ( 1 + UG_{in} )^2}$
We see that a negative $U$ generates a negative
$\mu_f$ which produces small hole pockets.  A negative $U$ also
enhance the spectral weight of unoccupied quasi-particles. This allows
us
to choose $U$ by requiring that the occupied spectral weight in $G_U$
is $1-x$. For general $u$ and $v$, we find
that the coherent part of $G'_U$ produces pocket-like
Fermi surfaces near
$\Sigma$ (see Fig 2) which are determined by
$2E_f(1+UG_{in})=Ux(u^2-v^2)$. The quasi-particle weight at the Fermi
surface
$Z = {2E^{2}_f\over xU^2}$ vanishes at $\Sigma$
and is very small
on the outer edge of the pocket,
making
it hard to detect.  As we approach the uRVB phase, $\Delta$ decreases
and the
pockets are elongated,
while their area increases with doping.
  Eventually the inner edges of the pockets join
together to form a large Fermi surface, with low lying excitations near
$M$ and a shape which resembles the experiment.\cite{1}

The incoherent background of Im$G_U$ contains two broad peaks separated
by
$2t$, shown in Fig 2a at $\Sigma$ and $M$.  This follows from the pseudo
gap
in the boson density of states and is a direct consequence of the
staggered flux.
Coherence factors cause a transfer of spectral weight from the
low energy to the high energy peak as one goes from $\Gamma$ to $\Sigma$
to
$X$,\cite{Lau}
in qualitative agreement with exact diagonalization results.\cite{15}
\begin{figure}
\epsfxsize=3.5truein
\centerline{\epsffile{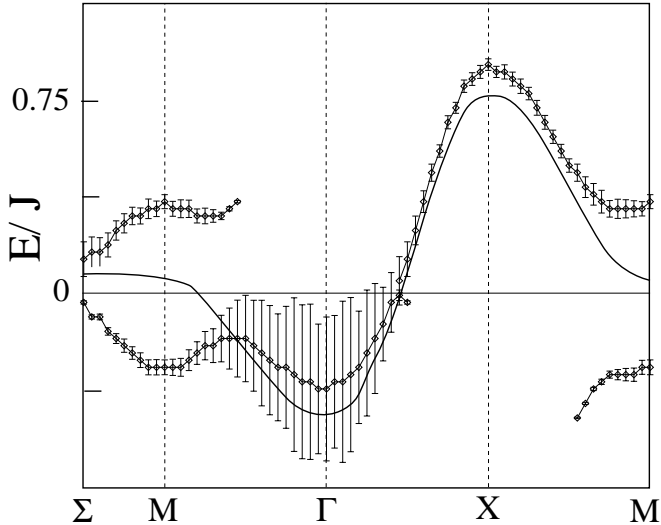}}
\caption{
The points describe the dispersion of the quasi-particle peaks for the
sF phase in Fig. 2. The vertical bars are proportional
to the peak values of Im$G_U$ which reflect the quasi-particle weight.
The solid curve is the quasi-particle dispersion for a uRVB phase
with the same doping concentration but at a higher temperature
($T/J=0.19$).
}
\label{fig3}
\end{figure}

Fig 3 shows the dispersion of the quasi-particle peaks and their
spectral
weights.  Comparing the sF and uRVB dispersions, we see clearly
 a splitting of the spectral weight along
$\Sigma$ to $M$ which is remarkable for  a translational
invariant state.
In our theory the splitting is naturally related to the spin gap.
  We also note that at half filling, the bottom of the
band
at $M$ and $\Gamma$ are degenerate.  As doping is increased,
the spin gap shrinks and
 the occupied
band near $M$ moves up in energy, eventually producing the flat band
near the Fermi surface seen in photo-emission experiments.  In our
calculation, the band at $\Gamma$ has been pushed up in energy from its
bare value near $-J$ to $-J/2$ due to the inclusion of $U$, whereas
the experimental value is closer to $-2J$.  We believe this feature, as
well
as the large spectral weight near $\Gamma$, is an artifact of our crude
treatment of the gauge fluctuation.

Fig 2b shows, in more detail, how the Fermi surface disappears in the
sF phase.  In scan C, a ghost band below the Fermi energy
is quite visible after the main peak
goes
above the Fermi surface.  This band is connected to the occupied band in
Fig 3 as the $M$ point is approached.
The apparence of the ghost band and the sudden reduction of the
quasiparticle spectral weight in the ghost band have been observed
in the insulating cuprates.\cite{12}
The inner edge of the Fermi pocket in the insert of Fig. 2
is determined from the position of the quasiparticle peaks at
$\omega=0$.
The quasiparticle peaks at the outer edge of the Fermi pocket
are not visible in our numerical result,
and the full ellipse of the Fermi pocket is completed
based on our analytic results on $G_U'$.

The $SU(2)$ MF theory shares many similar physical properties
with the
$U(1)$ MF theory (where the spin gap is generated
by the $d$-wave
pairing of the fermions).  However, there are some qualitative
distinctions.
1)  The $d$-wave state in the $U(1)$ theory does not
produce the
double-peak structure in the incoherent background of Im$G$.  One needs
to
use a flux phase in the $U(1)$ theory to produce the double-peak
 (at the expense of breaking translation or time reversal
symmetry\cite{Lau}).
2)  The $d$-wave state in the $U(1)$ theory does not have
Fermi
pockets at finite doping, even if we include the gauge interaction
as we did in the $SU(2)$ theory.
3)  The sF phase in the $SU(2)$ theory contains a gapless
$U(1)$ gauge
field, which is absent  in the corresponding
$d$-wave state in the $U(1)$ theory.
It has been pointed out that the existence of a mass gap in the $U(1)$
theory may de-stablize the $d$-wave state.\cite{UL2} The sF phase may be
more stable from this point of view.

\vskip 3mm
We would like to thank Z.X. Shen and M. Sigrist for very helpful
discussions.
PAL is supported by
NSF-MRSEC grant DMR-94-00334
and XGW is supported by NSF grant DMR-94-11574
and A.P. Sloan fellowship.


\end{multicols}
\end{document}